\documentclass[reprint,amsmath,amssymb,
 aps,bbl,showpacs,preprintnumbers,nofootinbib,
 longbibliography,longtable,
  lengthcheck]{revtex4}
\usepackage{bm}
\usepackage{graphicx,color}
\usepackage{epsfig,epstopdf}
\usepackage{amsfonts,amsmath,amssymb,bm,dsfont,eucal}
\usepackage[english]{babel}
\usepackage{hyperref}
\newcommand{\be}{\begin{equation}}
\newcommand{\ee}{\end{equation}}
\newcommand{\bea}{\begin{eqnarray}}
\newcommand{\eea}{\end{eqnarray}}
\newcommand{\nn}{\nonumber}

\begin{document}

\preprint{BARI-TH/625-10}
\title{$B_s\to f_0(980)$ form factors and $B_s$ decays into $f_0(980)$}
\author{ Pietro Colangelo,  Fulvia  De Fazio and Wei Wang}
\affiliation{Istituto Nazionale di Fisica Nucleare, Sezione di Bari, Italy}

\begin{abstract}
We compute the $B_s\to f_0(980)$ transition form factors using  light-cone
QCD sum rules at leading order in the strong coupling constant, and also   including an estimate of next-to-leading
order corrections. We use the  results to predict the branching fractions of the rare
decay modes $B_s \to f_0 \ell^+ \ell^-$ and $B_s \to f_0 \nu \bar \nu$, which turn out to be   ${\cal O}(10^{-7})$ ($B_s\to
f_0(980)\ell^+\ell^-$, with $\ell=e,\mu$), ${\cal O}(10^{-8})$ ($B_s\to
f_0(980)\tau^+\tau^-$) and ${\cal O}(10^{-6})$($B_s\to f_0(980)\nu\bar\nu$). We also predict the branching ratio of $B_s\to J/\psi
f_0(980)$ decay under the factorization assumption, and discuss the role of this channel for the determination of the $B_s$ mixing phase compared to the golden mode $B_s \to J/\psi \phi$. As a last application, we consider $D_s \to f_0$ form factors,  providing a determination of the branching ratio of $D_s \to f_0 e^+ \nu_e$.
\end{abstract}

\pacs{13.20.He, 13.20.Fc, 13.25.Hw} \maketitle

\section{Introduction}
Theoretical and experimental efforts aimed at disclosing physics beyond the standard model (SM) proceed in several directions. Among these, there is the study  of rare processes which are induced only at loop level in the SM and are therefore sensitive to new physics (NP) contributions which may potentially enhance their  small ($< 10^{-5}$) branching ratios \cite{Ball:2000ba}. Another testing ground is  the precise study of CP violation.
It has been realized that the amount of CP violation within the SM is too small to explain the observed baryon asymmetry of the Universe \cite{Sakharov:1967dj}, a conclusion confirmed by  recent analyses \cite{Huet:1994jb}. Since the only source of CP violation in the SM is the complex phase of the Cabibbo Kobayashi Maskawa (CKM) mixing matrix, the determination of the elements of this matrix and of their relative phases is of primary importance, in order to disentangle souces of additional contributions to CP violation.
 As well known, the task is afforded through the study of the so called unitarity triangles,  the graphical representations of the conditions stemming from  unitarity of the CKM matrix. The most studied triangle is the one which relates the CKM elements involved in $B$ decays. Direct and indirect determinations of its sides and angles  lead to a  picture of CP violation coherent with the SM description. 
Also in this case, investigation of effects  predicted to be small in the SM is a promising strategy to reveal new physics.

$B_s$ mesons provide the possibility  to search for  new physics 
scenarios exploiting both the strategies outlined above. On the one hand, rare $B_s$ decays induced by $b \to s$ transition  are suppressed in the SM, as with all decay modes governed by such a transition, and new physics effects may enhance their branching fractions.
For example,  it has been shown that, in presence of a single universal  extra dimension compactified on a circle with radius $R$, the rates of  $B_s \to \phi \nu \bar \nu$, $B_s \to \eta^{(\prime)} \ell^+ \ell^-$ and $B_s \to \eta^{(\prime)} \nu \bar \nu$
are enhanced when $R^{-1}$ decreases, while the opposite happens in the case of  $B_s \to \phi \gamma$, which has a smaller branching fraction with respect to  SM  for small values of $R^{-1}$
  \cite{Colangelo:2007jy}.

On the other hand, the analysis of the unitarity triangle of  CKM elements relevant for $B_s$ decays
is  an important test of the SM description of CP violation. The triangle is defined by the relation
\be V_{us}V_{ub}^*  + V_{cs}V_{cb}^* + V_{ts}V_{tb}^*  = 0 \label{bsUT}\,\,. \ee
One of its angles, $ \beta_s$, defined as $ \beta_s= Arg\left[ - \frac{V_{ts} V^*_{tb}}{V_{cs} V^*_{cb}}\right]$, is half of the phase of $B_s - {\bar B}_s$ mixing,  and is predicted  to be tiny in the SM: $\beta_s \simeq 0.019$ rad. Recent data obtained by the CDF~\cite{Aaltonen:2007he} and D0~\cite{:2008fj} Collaborations, based on  the angular analysis of  $B_s\to J/\psi \phi$,  indicate much larger values,  although with sizable uncertainties, so that the precise measurement of $\beta_s$ represents one of the priorities in the physics programs at the hadron colliders and at the  $B$ factories operating at the $\Upsilon(5S)$ peak  \cite{machines}.

In this paper we consider $B_s$ decays in both respects. We  compute the $B_s \to f_0(980)$~\footnote{Hereafter, we 
use $f_0$ to denote the $f_0(980)$ meson.} form factors using  light-cone QCD sum rules (LCSR) at the  leading order in the strong coupling constant (Sect. \ref{sec:formfactors-LCSR},\ref{sec:results-discussions}) and including an estimate  of next-to-leading (NLO) corrections (Sect. \ref{sec:radiative-corrections}). In Sect. \ref{PA-A}, we
use the results to predict the branching fractions of the rare decay modes $B_s\to f_0 \ell^+ \ell^-$ and $B_s\to f_0\nu\bar\nu$ in the SM.
The  form factors are also a necessary ingredient to study the nonleptonic mode  $B_s\to J/\psi
f_0$ which,  together with $B_s\to J/\psi \phi$, permits one to access the phase $\beta_s$ \cite{Stone:2008ak}. Our predictions for this mode are collected in Sect. \ref{PA-B}.
As a byproduct of the calculation, we  explore
the $D_s\to f_0e^+\nu_e$ decay channel, the branching ratio of which  has been recently
measured by the CLEO Collaboration~\cite{:2009cm,:2009fi}.
Conclusions are presented in the last section.
\vfill

\section{Light-Cone QCD Sum Rule calculation of  $B_s\to f_0$ form factors \label{sec:formfactors-LCSR}}

The matrix elements involved in $B_s\to f_0$ transitions can be parameterized  in terms of  form factors as
\begin{eqnarray}
&&\langle f_0(p_{f_0})|\bar s \gamma_\mu\gamma_5 b |\overline {B}_s(p_{B_s})\rangle =  \nonumber \\ 
&&-i \Big\{F_1(q^2)\Big[P_\mu
 -\frac{m_{B_s}^2-m_{f_0}^2}{q^2}q_\mu\Big] + F_0(q^2)\frac{m_{B_s}^2-m_{f_0}^2}{q^2}q_\mu\Big\}, \,\,\, \nonumber \\
 \label{F1-F0} 
 \eea
 \bea
&&\langle {f_0}(p_{f_0})|\bar s\sigma_{\mu\nu}\gamma_5q^\nu b |\overline {B}_s(p_{B_s})\rangle = \nonumber \\
 &&-\frac{F_T(q^2)}{m_{B_s}+m_{f_0}}   \Big[q^2P_\mu
 -(m_{B_s}^2-m_{f_0}^2)q_\mu\Big], \label{FT}
\end{eqnarray}
where $P=p_{B_s}+p_{f_0}$ and $q=p_{B_s}-p_{f_0}$. In this section
we describe the calculation of the three functions $F_1$, $F_0$ and
$F_T$  using the method of light-cone QCD sum rules. For the sake of the calculation, it is
 convenient to define the auxiliary
form factors $f_+$ and $f_-$,
\bea
 \langle {f_0}(p_{f_0})|\bar s\gamma_\mu\gamma_5 b |\overline {B}_s(p_{B_s})\rangle& =&
 -i\left\{f_+(q^2) P_\mu +f_-(q^2)q_\mu\right\} \nonumber \\
\eea
in terms of which $F_1$ and $F_0$ read:
\begin{eqnarray}
 F_1(q^2)&=& f_+(q^2) \,\, , \nonumber \\
 F_0(q^2)&=&f_+(q^2)+\frac{q^2}{m_{B_s}^2-m_{f_0}^2} f_-(q^2) \,\, .
\end{eqnarray}
As a reconciliation of the original QCD sum rule approach \cite{SVZ}  and the
application of perturbation theory to hard processes,  LCSR  \cite{braun}
present several advantages in the calculation of quantities such as the  heavy-to-light meson form factors. The method
includes both hard scattering and  soft contributions. In the
hard scattering region the operator product expansion (OPE) near
the light-cone is applicable. Based on the light-cone OPE, hadronic
quantities, like  form factors, are expressed
as a convolution of  light-cone distribution amplitudes (LCDA)
with a perturbatively calculable hard kernel. The leading
twist  and a few subleading twist LCDA give the dominant
contribution, while higher twist terms are power suppressed. The
LCSR approach has been successfully applied to compute the hadronic parameters involved in many different processes   \cite{Colangelo:2000dp}.

The starting point for a LCRS evaluation of  form factors is the
correlation function of suitably chosen quark currents.  Here  we consider the correlation function
\begin{eqnarray}
 \Pi(p_{f_0},q)&=& i \int d^4x \, e^{iq\cdot x} \langle {f_0}(p_{f_0})|{\rm
 T}\left\{j_{\Gamma_1}(x),j_{\Gamma_2}(0)\right\}|0\rangle \,\,\,\, \nonumber \\
 \label{corr}
\end{eqnarray}
where $j_{\Gamma_1}$ is one of the currents appearing 
in the  matrix elements (\ref{F1-F0}-\ref{FT})  defining
the form factors: $j_{\Gamma_1}=J_\mu^5=\bar s\gamma_\mu\gamma_5b$
for $F_1$ and $F_0$, and $j_{\Gamma_1}=J_\mu^{5T}=\bar
s\sigma_{\mu\nu}\gamma_5 q^\nu b$ for $F_T$.  The
 current  $j_{\Gamma_2}$  interpolates  the $B_s$ meson:  we choose
$j_{\Gamma_2}=\bar b i\gamma_5 s$. Its matrix element between the vacuum and $B_s$ is given in terms of the decay 
constant $f_{B_s}$,
\begin{eqnarray}
 \langle \overline B_s(p_{B_s})| \bar b i\gamma_5 s|0\rangle &=&
 \frac{m_{B_s}^2}{m_{b}+m_s}f_{B_s}.
\end{eqnarray}
We also introduce  the
$f_0(980)$ decay constant ${\bar f}_{f_0}$,
\begin{eqnarray}
 \langle {f_0}(p_{f_0})|\bar s  s|0\rangle =m_{f_0}\bar  f_{f_0}
 \label{fbarf0}
\end{eqnarray}
 needed in the following;
${\bar f}_{f_0}$ has been evaluated by several
groups~\cite{DeFazio:2001uc,Bediaga:2003zh,Cheng:2005ye,Cheng:2005nb}.

The LCSR method consists in evaluating the correlation
function  Eq.(\ref{corr}) both at the hadronic  level and in QCD. Equating
the two  representations provides one with a sum rule suitable  to derive  the form
factors.

The hadronic representation of the correlation function in
(\ref{corr}) 
\begin{eqnarray}
 & &\Pi^{\rm HAD}(p_{f_0},q)= \nn \\ &&  \frac{\langle {f_0}(p_{f_0})|j_{\Gamma_1}|\overline {B}_s(p_{f_0}+q)\rangle \langle \overline {B}_s(p_{f_0}+q)|j_{\Gamma_2}|0\rangle}
 {m_{B_s}^2-(p_{f_0}+q)^2}\nonumber \\
 &&+\sum_h \frac{\langle {f_0}(p_{f_0})|j_{\Gamma_1}|h(p_{f_0}+q)\rangle \langle h(p_{f_0}+q)|j_{\Gamma_2}|0\rangle}
 {m_h^2-(p_{f_0}+q)^2} \nonumber \\
\end{eqnarray}
consists in  the contribution of the $\bar B_s$ meson and of the
higher resonances and the continuum of states $h$.
 In a one-resonance+continuum  representation, the correlation function
can be written as
\begin{eqnarray}
 &&\Pi^{\rm HAD}(p_{f_0},q)=\nn \\& &  \frac{\langle {f_0}(p_{f_0})|j_{\Gamma_1}|\overline
 {B}_s( p_{f_0}+q)\rangle \langle \overline {B}_s(p_{f_0}+q)|j_{\Gamma_2}|0\rangle}
 {m_{B_s}^2-(p_{f_0}+q)^2}\nonumber \\&&+ \int_{s_0}^\infty ds \frac{\rho^h(s,q^2)}
 {s-(p_{f_0}+q)^2}, \label{hadronic}
\end{eqnarray}
where  higher resonances and  the continuum of states are
described in terms of the spectral function $\rho^h(s,q^2)$, which
 contributes starting from a threshold $s_0$.  
 
 At the quark level, the correlation function can be
 evaluated in QCD in the deep Euclidean region, where  it  can be written as
\begin{eqnarray}
 \Pi^{\rm QCD}(p_{f_0},q)&=&   \frac{1}{\pi}\int_{(m_b+m_s)^2}^\infty ds \, \frac{{\rm Im}\Pi^{\rm QCD}(s,q^2)}
 {s-(p_{f_0}+q)^2} \,. \,\,\,\,\, \label{QCD-repr}
\end{eqnarray}
Using  global quark-hadron duality,  the integral  in
(\ref{hadronic}) can be identified with the corresponding quantity
 in the QCD representation (\ref{QCD-repr}):
\begin{eqnarray}
\int_{s_0}^\infty  ds {\rho^h(s,q^2) \over s-(p_{f_0}+q)^2}&=&\frac{1}{\pi}\int_{s_0}^\infty ds {{\rm Im}\Pi^{\rm QCD}(s,q^2)\over s-(p_{f_0}+q)^2}\,. \nonumber \\ 
\end{eqnarray}
A Borel transformation of the hadronic and of the QCD  expressions of the
correlation function is  carried out,   defined as:
\bea {\cal B} [{\cal
F}(Q^2)]=lim_{Q^2 \to \infty, \; n \to \infty, \; {Q^2 \over
n}=M^2}\; \nonumber \\{1 \over (n-1)!} (-Q^2)^n \left({d \over dQ^2} \right)^n
{\cal F}(Q^2) \; , \label{tborel} \eea  
where ${\cal F}$  is a  function of $Q^2=-q^2$ and $M^2$ is  the Borel parameter, so that  
 \be {\cal B} \left[ { 1 \over (s+Q^2)^n }
\right]={\exp(-s/M^2) \over (M^2)^n\ (n-1)!} \; . \label{bor} \ee
This operation improves the convergence of the OPE series  by
factorials of $n$ and, for suitably chosen values of $M^2$,
enhances the contribution of the low lying states to the correlation function.

Applying the transformation to both $\Pi^{\rm HAD}$ and $\Pi^{\rm QCD}$ we
obtain the  sum rule
\begin{eqnarray}
  &&  {\langle {f_0}(p_{f_0})|j_{\Gamma_1}|\overline {B}_s(p_{B_s})\rangle \langle \overline {B}_s(p_{B_s})|j_{\Gamma_2}|0\rangle}
 {\rm
 exp}\left[-\frac{m_{B_s}^2}{M^2}\right]=\nonumber \\
 &&\frac{1}{\pi}\int_{(m_b+m_s)^2}^{s_0}ds \,
 {\rm exp}[-s/M^2] \, \, {\rm Im}\Pi^{\rm QCD}(s,q^2), \label{SR-generic}
\end{eqnarray}
where $p_{B_s}=p_{f_0}+q$. Eq.(\ref{SR-generic})  allows to
derive the sum rules for $f_+$, $f_-$ and $F_T$,  choosing
either  the current $j_{\Gamma_1}=J_\mu^5$ or the current  $j_{\Gamma_1}=J_\mu^{5T}$.

The calculation of $\Pi^{\rm QCD}$ is based on the expansion of the
T-product in (\ref{corr}) near the light-cone, which  produces
matrix elements of non-local quark-gluon operators. 
In the description of $f_0$ as a $s \bar s$ state modified by some hadronic dressing \cite{DeFazio:2001uc},
these can be defined in
terms of  $f_0$ light-cone distribution amplitudes  of
increasing twist:
\begin{eqnarray}
 \langle {f_0}(p_{f_0})|\bar s (x)\gamma_\mu s(0)|0\rangle &=& \bar f_{f_0}
 p_{{f_0}\mu}
 \int_0^1 due^{i up_{f_0}\cdot x}\Phi_{f_0}(u),\nonumber\\
 \langle {f_0}(p_{f_0})|\bar s (x)  s(0)|0\rangle &=& m_{f_0} \bar f_{f_0}
 \int_0^1 du e^{iup_{f_0}\cdot x}\Phi_{f_0}^s(u),\nonumber \\
 \langle {f_0}(p_{f_0})|\bar s(x)\sigma_{\mu\nu}   s(0)|0\rangle &=&
 - \frac{m_{f_0}}{6}\bar f_{f_0} (p_{{f_0}\mu} x_\nu -p_{{f_0}\nu} x_\mu) \nonumber \\
 &\times&\int_0^1 du e^{i up_{f_0}\cdot x} {\Phi_{f_0}^\sigma(u)} ,
\end{eqnarray}
where the  LCDA $\Phi_{f_0}$ is twist-2, and   the other two
 are twist-3, and are normalized  as
\be
 \int_0^1 du \Phi_{f_0}(u)=0,\;\;\; \int_0^1 du \Phi_{f_0}^s(u)=\int_0^1
 du\Phi_{f_0}^\sigma(u)=1.
\ee

In terms of these LCDA, the sum rules for the three form
factors read:
\begin{widetext}
\begin{eqnarray}
  f_+(q^2)&=& \frac{m_b+m_s}{2m_{B_s}^2f_{B_s}}\bar f_{f_0}{\rm exp}\left[\frac{m_{B_s}^2}{M^2}\right]
  \left\{\int_{u_0}^1\frac{du}{u}{\rm exp}\left[-\frac{m_b^2+u\bar u m_{f_0}^2-\bar uq^2}{uM^2}\right]
 \left[-m_b\Phi_{f_0}(u)+um_{f_0}\Phi_{f_0}^s(u)+\frac{1}{3}m_{f_0}\Phi_{f_0}^\sigma(u)\right.\right.\nonumber\\
 &&\;\;\;\left.\left. +\frac{
 m_b^2+q^2-u^2m_{f_0}^2}{uM^2}\frac{m_{f_0}\Phi_{f_0}^\sigma(u)}{6}
 \right]
 +\exp{[-s_0/M^2]}\frac{m_{f_0}\Phi_{f_0}^\sigma(u_0)}{6}\frac{m_b^2-u_0^2m_{f_0}^2+q^2}
 {m_b^2+u_0^2m_{f_0}^2-q^2}\right\},\label{eq:fplus}\\
 f_-(q^2)&=& \frac{m_b+m_s}{2m_{B_s}^2f_{B_s}}\bar f_{f_0}{\rm
 exp}\left[\frac{m_{B_s}^2}{M^2}\right]\left\{\int_{u_0}^1\frac{du}{u}{\rm
 exp}\left[-\frac{m_b^2+u\bar u m_{f_0}^2-\bar uq^2}{uM^2}\right]
 \left[ m_b\Phi_{f_0}(u)+(2-u) m_{f_0}\Phi_{f_0}^s(u)\right.\right.\nonumber\\
 &&\;\;\;\left.\left.+\frac{1-u}{3u}m_{f_0}\Phi_{f_0}^\sigma(u) -\frac{u({m_b^2+q^2-u^2m_{f_0}^2})+2(
 m_b^2-q^2+u^2m_{f_0}^2)}{u^2M^2}\frac{m_{f_0}\Phi_{f_0}^\sigma(u)}{6}
 \right]\right.\nonumber\\
 &&\left. -\frac{ u_0({m_b^2+q^2-u_0^2m_{f_0}^2})+2(
 m_b^2-q^2+u_0^2m_{f_0}^2) }{u_0(m_b^2+u_0^2m_{f_0}^2-q^2)}
  \exp{[-s_0/M^2]}\frac{m_{f_0}\Phi_{f_0}^\sigma(u_0)}{6}\right\},\label{eq:fminus}\\
 F_T(q^2)&=& \frac{(m_{B_s}+m_{f_0})(m_b+m_s)}{m_{B_s}^2f_{B_s}}\bar f_{f_0}{\rm exp}
 \left[\frac{m_{B_s}^2}{M^2}\right]
 \left\{\int_{u_0}^1\frac{du}{u} {\rm exp}\left[-\frac{(m_b^2-\bar uq^2+u\bar
 um_{f_0}^2)}{uM^2}\right]
 \left[-\frac{\Phi_{f_0}(u)}{2}+m_b\frac{m_{f_0}\Phi_{f_0}^\sigma(u)}{6uM^2}\right]\right.\nonumber\\
 &&\;\;\;\left.+m_b\frac{m_{f_0}\Phi_{f_0}^\sigma(u_0)}{6}
   \frac{\exp[-s_0/M^2]}{m_b^2-q^2+u_0^2m_{f_0}^2}\right\},\label{eq:ftensor}
\end{eqnarray}
where
\begin{eqnarray}
 u_0&=&\frac{m_{f_0}^2+q^2-s_0+\sqrt{(m_{f_0}^2+q^2-s_0)^2+4m_{f_0}^2(m_b^2-q^2)}}{2m_{f_0}^2}.\label{eq:u0}
\end{eqnarray}\end{widetext}
Our formulae can be compared  to the
ones for the $B$-to-scalar meson form factors  in
Ref.~\cite{Wang:2008da}, where the case of the meson $a_0$
is considered.  We find  differences in the expression of the form
factor $f_+$.

\section{Numerical results and discussions}
\subsection{Leading order results}\label{sec:results-discussions}

Based on the conformal spin invariance, the LCDA can be expanded in terms of Gegenbauer polynomials $C_n^{3/2}$. The expansion
of the twist-2 LCDA  $\Phi_{f_0}(u)$  reads:
\be
 \Phi_{f_0}(u)=  6u(1-u)\left\{B_0+\sum_{n=1} B_n
 C_n^{3/2}(2u-1)\right\}\,\,. \label{phif0}
\ee
Due to the charge conjugation invariance, all even Gegenbauer moments of  $\Phi_{f_0}(u)$ vanish, so that  $B_{2 m}=0$ for $m=0,1,\cdots$ in (\ref{phif0});  as for the odd moments, we
include only the first one, using the value of the coefficient $B_1=-0.78\pm0.08$ fixed in 
ref.\cite{Cheng:2005nb}. For the twist-3 LCDA, due to the lack of
knowledge about their  moments, we use the asymptotic form, i.e. the first term of the Gegenbauer expansion, 
\begin{eqnarray}
 \Phi_{f_0}^s(u)&=&1,\;\;\;\;\;\; \Phi_{f_0}^\sigma(u)=
 6u(1-u).
\end{eqnarray}

Let us quote the numerical values of the other physical parameters. The meson masses are fixed to the PDG values $m_{B_s} = 5.366\,
{\rm GeV}$ and $m_{f_0}=0.98\, {\rm GeV}$ \cite{Amsler:2008zzb},
while for quark masses we use $m_b=4.8\, {\rm GeV}$ and $m_s=0.14\,
{\rm GeV}$ \cite{Amsler:2008zzb,quarkmass}. As for the decay constants, we use
 $f_{B_s}=(0.231\pm 0.015)\, {\rm GeV}$
\cite{Gamiz:2009ku} and
 $\bar f_{f_0}= (0.18\pm0.015)\, {\rm GeV}$
 \cite{DeFazio:2001uc}\footnote{In ref.~\cite{Cheng:2005nb}  
  a  larger result is reported:
$ \bar f_{f_0}=(0.37\pm0.02)\,  {\rm GeV}$.}. The
threshold  $s_0$ is fixed to $s_0=(34\pm 2)\, {\rm
 GeV}^2$, which should correspond to the mass squared of the first radial excitation of $B_s$.
\begin{figure}[b]
\includegraphics[width=0.4\textwidth]{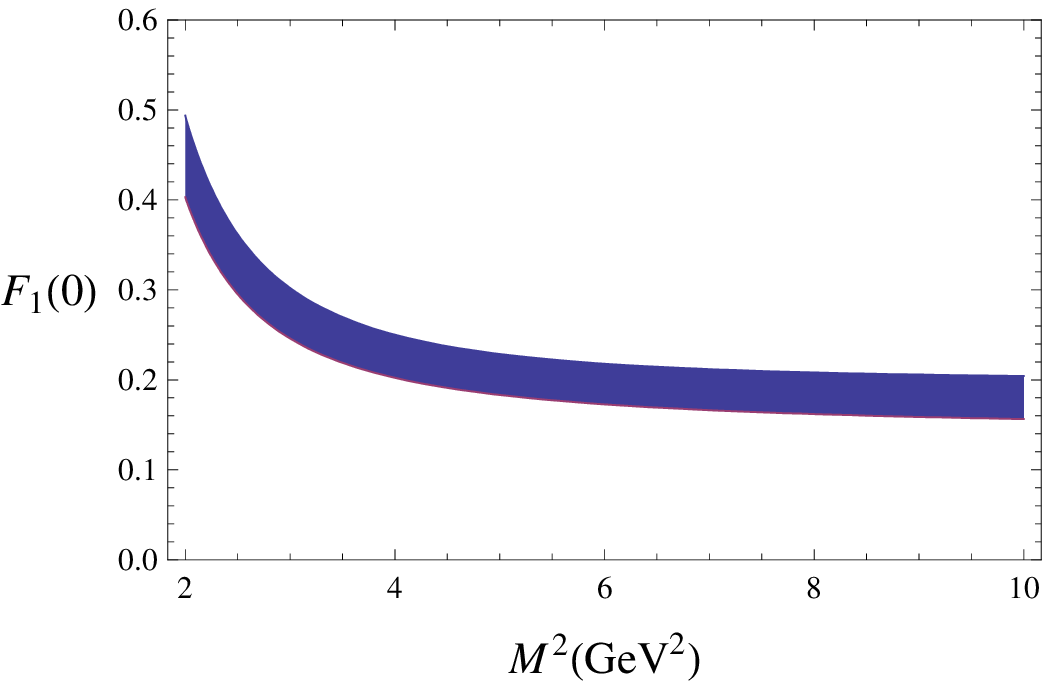}
\includegraphics[width=0.4\textwidth]{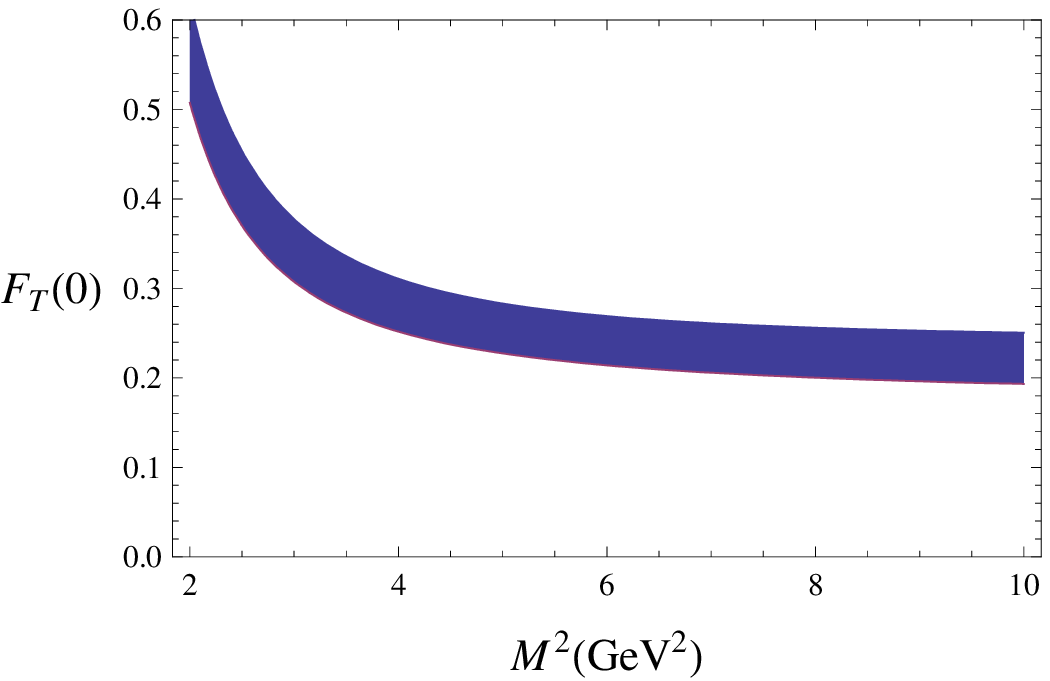}
\caption{Dependence on the Borel parameter $M^2$ of the $B_s \to f_0$ form factors at $q^2=0$:  $F_1(0)=F_0(0)$ (upper panel) and
$F_T(0)$ (lower panel). }\label{fig:M2dependence}
\end{figure}

With these numerical inputs, the sum rules
(\ref{eq:fplus})-(\ref{eq:ftensor}) provide us with the form factors for
each value of $q^2$ as a  function of the Borel parameter. The
result is obtained requiring stability against variations of
$M^2$.

\begin{figure}[t]
\includegraphics[width=0.4\textwidth]{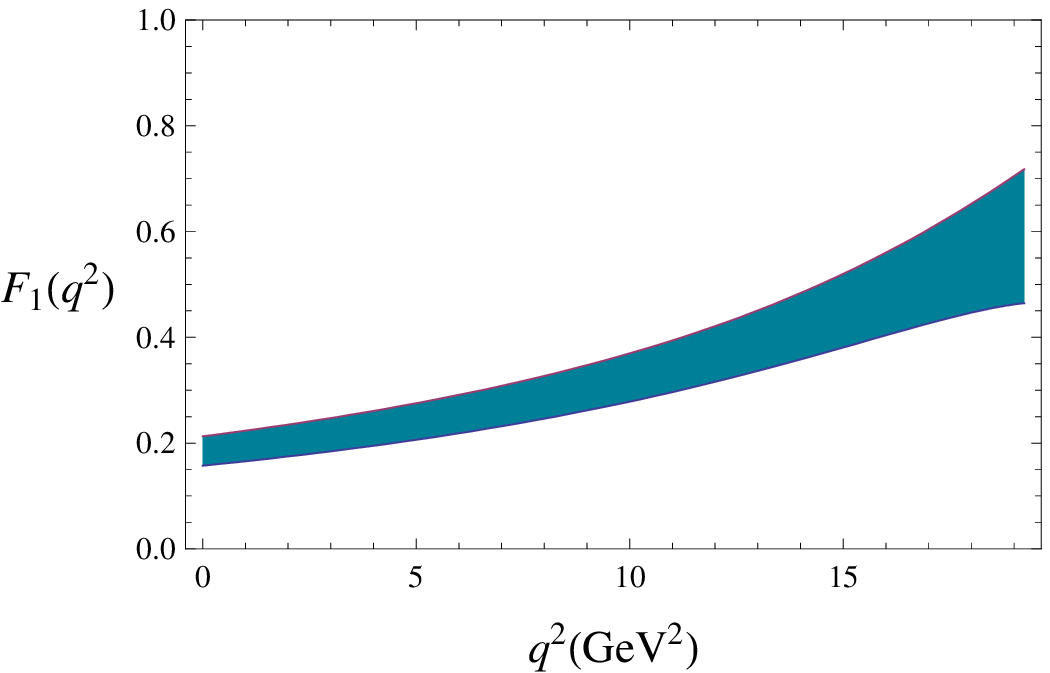}
\includegraphics[width=0.4\textwidth]{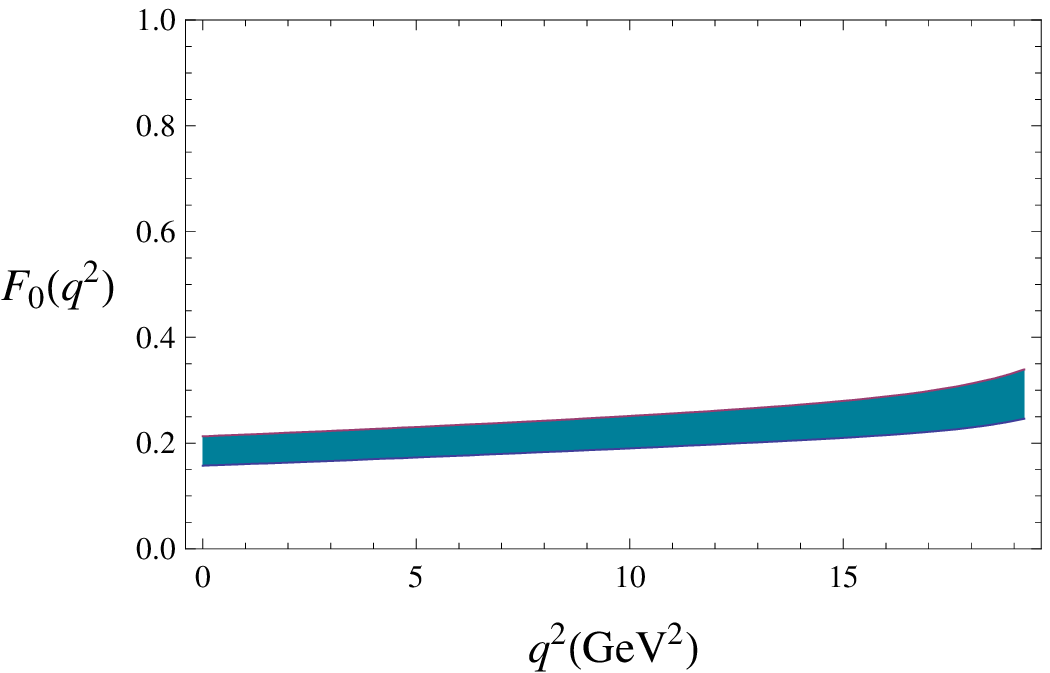}\\
\includegraphics[width=0.4\textwidth]{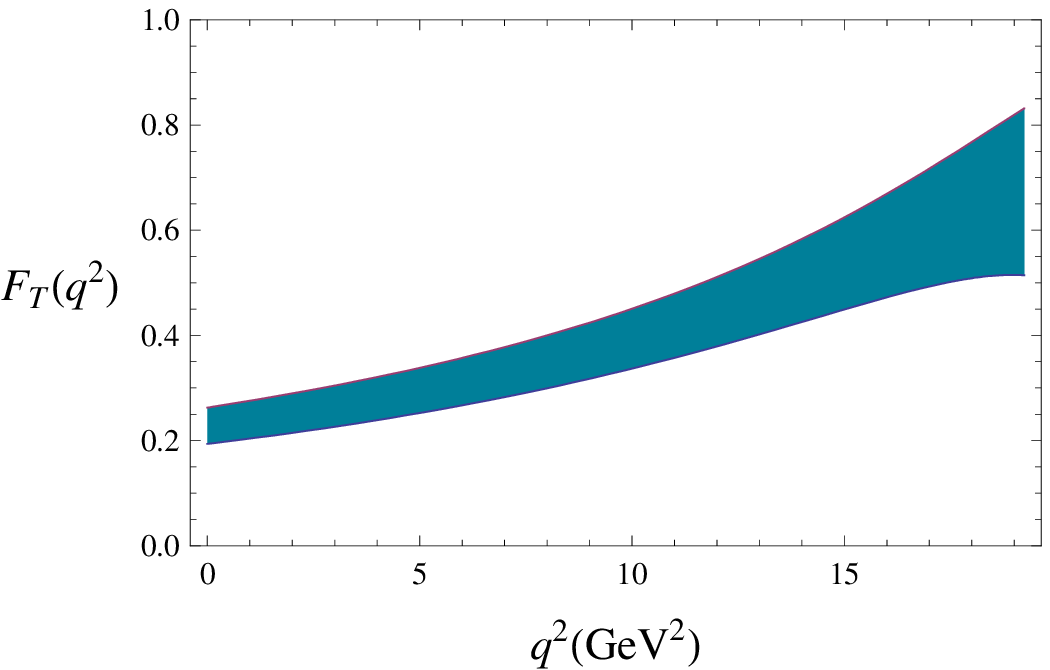}
 \caption{$q^2$ dependence of the $B_s\to
f_0$ form factors. }\label{fig:LO-formfactor}
\end{figure}

In Fig.~\ref{fig:M2dependence}  we show the dependence of the form factors at $q^2=0$ on the Borel parameter
$M^2$. We  observe stability when
$M^2>6$ ${\rm GeV}^2$, and we fix $M^2=(8\pm2)\,  {\rm GeV}^2$.
To describe the form factors in the whole kinematically accessible
$q^2$ region, we adopt the  parameterization
\begin{eqnarray}
 F_i(q^2)&=&\frac{F_i(0)}{1-a_iq^2/m_{B_s}^2+b_i(q^2/m_{B_s}^2)^2}\,\,,\label{eq:formfactors-fit-form}
\end{eqnarray}
where $F_i$ denotes any function among $F_{1,0,T}$. The
parameters $a_i,b_i$ are obtained through fitting the form
factors in the small $q^2$ region (we choose  $0<q^2<15$
$ {\rm GeV}^2$);  the  results for  $F_i(0)$, $a_i$ and $b_i$
 are collected in
Table~\ref{table:LO-formfactor}, and the  $q^2$ dependence is depicted in Fig.~\ref{fig:LO-formfactor}.
 The uncertainties in the results reflect those  of the input parameters.   In
Table~\ref{table:LO-formfactor} we also report the values of the
form factors at zero-recoil  ($q^2_{\rm max}$) which are
derived using the  expression  in
Eq.~(\ref{eq:formfactors-fit-form}).

The results in Table~\ref{table:LO-formfactor} show
that the  parameters $a_i$ and $b_i$ determining the $q^2$ dependence
 are close to each other in the case of  $F_1$ and $F_T$. The reason is the following.
 In the heavy-quark limit and in the large energy  (LE) limit of the recoiled meson, the
three $B_s\to f_0$ form factors can be related to a single
universal function $\xi_{f_0}$ which is specific for $f_0$  and
does not depend on the Dirac structure of the current appearing in
the definition of the various matrix elements,  such as those in
Eqs. (\ref{F1-F0}-\ref{FT}) \cite{Charles:1998dr}. When the energy $E$  of the light
meson in the the final state  is
large, such  relations read as
\begin{eqnarray}
 \frac{m_{B_s}}{m_{B_s}+m_{f_0}}F_T(q^2)=F_1(q^2)=\frac{m_{B_s}}{2E}
 F_0(q^2),\label{eq:large-energy-limit}
\end{eqnarray}
where, neglecting $m_{f_0}^2$ but keeping $m_{f_0}$ in the
kinematical factors,  $E$ is related to $q^2$:
\begin{eqnarray}
 q^2=m_{B_s}^2-2m_{B_s} E.
\end{eqnarray}

\begin{table}
\caption{Parameters of the $B_s \to f_0$  form factors by LCSR at the leading order.
 The values of $F_i(q^2_{\rm max})$ are evaluated through Eq.(\ref{eq:formfactors-fit-form}). }\label{table:LO-formfactor}
\begin{tabular}{ c c c c c c}
\hline\hline & $F_i(q^2=0)$  & $a_i$ & $b_i$ & $F_i(q^2_{\rm max})$
\\\hline
 $F_1 $   &   $0.185\pm0.029$
   &   $1.44^{+0.13}_{-0.09}$  &   $0.59^{+ 0.07}_{-0.05}$&   $0.614^{+0.158}_{-0.102}$ \\
 $F_0 $   &   $0.185\pm0.029$
    &   $0.47^{+0.12}_{-0.09}$  &   $0.01^{+  0.08}_{-0.09}$&   $0.268^{+0.055}_{-0.038}$\\
 $F_T $   &   $0.228\pm0.036$
    &   $1.42^{+0.13}_{-0.10}$  &   $0.60^{+  0.06}_{-0.05}$&   $  0.714^{+0.197}_{-0.126}$ \\
\hline\hline
\end{tabular}
\end{table}

The first equality in Eq.~(\ref{eq:large-energy-limit}) shows that the large energy limit  predicts that
$F_1$ and $F_T$ have the same $q^2$ dependence. For the shape
parameters of $F_0$,  one can  obtain two relations through
the second equality:
\begin{eqnarray}
 a_0 &=&-1+a_1,\;\;\; b_0 = 1-a_1+b_1. \label{scet}
\end{eqnarray}
Using the results for $a_1$ and $b_1$, we find from (\ref{scet}):
$a_0^{(LE)}\simeq 0.44 \pm 0.1$ and $b_0^{(LE)}\simeq 0.15 \pm
0.12$;  therefore,
 the first relation in (\ref{scet}) is well
respected in our calculation,  while  not much can be said about the
second relation due to the  uncertainty affecting  $b_0$.

\subsection{Estimate  of the next-to-leading order corrections}\label{sec:radiative-corrections}

In order to provide an estimate of next-to-leading order effects in the determination of the $B_s\to f_0$ form factors, it is worth comparing this case to the calculation of   $B\to \pi$ form factors.
 In $B\to \pi$ transition, both the light quarks and the light $\pi$
meson have small masses which can be safely neglected, while the
strange quark and the scalar meson $f_0$  masses may
induce sizable effects.  Another observation is that, neglecting the quark masses, the 
Lorentz structures of pion and $f_0$ matrix elements differ by  a
minus sign in terms proportional to the twist-2 LCDA. Finally,
contributions from the twist-3 LCDAs in $B\to \pi$ transition are
characterized by the chiral scale parameter $\mu_\pi$, while 
 in $B_s\to f_0$ they are  proportional to the mass of $f_0$.

In LCSR, NLO corrections to $B\to \pi$ form
factors have been  studied by two
groups~\cite{Ball:2004ye,Duplancic:2008ix}, while the complete
expressions for the NLO corrections to $B_s\to f_0$ form factors are not known
 at present.   The expressions relevant 
for $B\to \pi$ form factors given in Ref.~\cite{Duplancic:2008ix} can be used to 
estimate the radiative corrections in the case $B_s\to f_0$,  keeping in mind  the three differences above. We
first consider the changes to the leading order result due to the
different treatment of quark  and hadron masses. Setting the
quark mass $m_s$ to zero, the values of the form factors are reduced by about
 $3\%$. The mass of $f_0$, the analogous of the pion mass
$m_\pi$ and the chiral scale parameter $\mu_\pi$, cannot be
put to  zero, as this would smear all terms
from twist-3 LCDA:  we set the mass square of $f_0$ to be zero
keeping the linear terms in the form factors,  obtaining an enhancement of the
form factors by about  $3\%$. After that, evolving all the scale-dependent  parameters to a scale of about  the Borel mass, $\mu \simeq 3$
GeV, we find that the leading order contributions are furtherly
enhanced, obtaining the central values: 
$F_1(0)=F_0(0)=0.216$, $a_1=1.50$,  $b_1=0.58$,    $ a_0=0.216$, $b_0=0.53$ and
$F_T(0)=0.262$, $a_t=1.46$,  $b_t=0.58$.
Then, radiative corrections to twist-2 and twist-3 LCDA are  also found  to be rather
small,  the $B_s\to f_0$ form factors being changed to 
 $F_1(0)=F_0(0)=0.238$ and $ F_T(0)=0.308$. The resulting values, with the inclusion the uncertainty due to the input parameters, are
 collected in Table \ref{tab:formfactors-mass-zero}; they are also used  in the phenomenological analysis, keeping in mind, however, that 
 the procedure used in their determination must be considered  as only approximate.
 
\begin{table}
\caption{ ${  B_s \to  f_0(980)}$ transition form factors obtained including an estimate of  next-to-leading order corrections (see text).
}\label{tab:formfactors-mass-zero}
\begin{tabular}{c c c c c }
\hline\hline 
& $F_i(q^2=0)$  & $a_i$ & $b_i$   \\\hline 
$F_1 $   &   $0.238\pm0.036$      &   $1.50^{+0.13}_{-0.09}$  & $0.58^{+0.09}_{-0.07}$ \\
$F_0 $   &   $ 0.238\pm 0.036$&$0.53^{+0.14}_{-0.10}$  &   $-0.36^{+ 0.09}_{-0.08}$    \\
$F_T$   &   $ 0.308\pm0.049$&   $1.46^{+0.14}_{-0.10}$  & $0.58^{+ 0.09}_{-0.07}$    \\
\hline\hline
\end{tabular} \end{table}
%

Before closing this section, it is worth mentioning that
the $B_s\to f_0(980)$ form factors have been computed by other  approaches: the method based on covariant light-front dynamics (CLFD) and
dispersion relation (DR)~\cite{ElBennich:2008xy},  the perturbative
QCD approach (PQCD)~\cite{Li:2008tk},   short-distance  QCD sum rules
(QCDSR)~\cite{Ghahramany:2009zz}.  The results are collected in
Table~\ref{table:formfactor-comparison}. The form
factors by  PQCD are proportional to the $f_0$ decay constant, while those  by short-distance QCD sum rules are
proportional to the inverse of this constant. Thus,  a larger
decay constant,  $\bar f_{f_0}=0.37$ GeV as reported and used in \cite{Cheng:2005nb},  gives
larger form factors in the PQCD approach  and smaller ones in QCDSR with respect to ours.
Taking into account the difference in the decay
constant,  the results  in Refs.~\cite{Li:2008tk,Ghahramany:2009zz} are
consistent with ours, while the two  results  in
Ref.~\cite{ElBennich:2008xy} are sensibly larger.

\begin{table}
\caption{$B_s\to f_0(980)$ form factors at $q^2=0$. Results
evaluated by CLFD/DR~\cite{ElBennich:2008xy},
PQCD~\cite{Li:2008tk} and QCDSR~\cite{Ghahramany:2009zz} approaches are collected
for  a comparison. }\label{table:formfactor-comparison}
\begin{tabular}{c cccccc}
 \hline \hline &  CLFD/DR &  PQCD & QCDSR & This work \\\hline
  $F_1(0)$ & 0.40/0.29~\footnote{using  $f_{B_s}=0.259$ GeV}& $0.35_{-0.07}^{+0.09}$~\footnote{using  $\bar f_{f_0}=0.37$ GeV} & $0.12\pm0.03$~\footnote{using $\bar f_{f_0}=0.37$ GeV and $f_{B_s}=0.209$ GeV.}  &$0.185\pm0.029$\\
  $F_T(0)$&  & $0.40_{-0.08}^{+0.10}$ $^b$ &$-0.08\pm0.02$ $^c$  &$0.228\pm0.036$\\
 \hline \hline
\end{tabular}
\end{table}

\section{Phenomenological applications}

\subsection{Semileptonic $\bar B_s\to f_0 \ell^+ \ell^-$ and $\bar B_s\to f_0
\nu\bar\nu$ decays}\label{PA-A}

As a first application of our study,  we predict the branching ratios of the  decays $\bar B_s\to f_0\ell^+\ell^-$ and $\bar B_s\to f_0 \nu\bar\nu$,  processes which,  being induced by the flavor-changing neutral current transition $b \to s$,   are potentially important  for detecting new physics effects.

The SM  $ \Delta B =1$, $\Delta S =-1$ effective Hamiltonian describing  the  transition $b \to s \ell^+
\ell^-$ can be expressed in terms of a set of local operators:
\begin{equation}
H_{b \to s \ell^+ \ell^-}\,=-\,4\,{G_F \over \sqrt{2}} V_{tb}
V_{ts}^* \sum_{i=1}^{10} C_i(\mu) O_i(\mu) \,\,\,\, , \label{hamil}
\end{equation}
\noindent  $G_F=1.166\times 10^{-5}{\rm GeV}^{-2}$  being the Fermi constant and $V_{ij}$ the
elements of the CKM mixing matrix
(since the ratio
$\displaystyle \left |{V_{ub}V_{us}^* \over V_{tb}V_{ts}^*}\right
|$ is ${\cal O}(10^{-2})$, we neglect terms proportional to $V_{ub} V_{us}^*$). The operators $O_i$ are written in
terms of quark and gluon fields:
\begin{eqnarray}
O_1&=&({\bar s}_{L \alpha} \gamma^\mu b_{L \alpha})
      ({\bar c}_{L \beta} \gamma_\mu c_{L \beta}) \,,\nonumber \\
O_2&=&({\bar s}_{L \alpha} \gamma^\mu b_{L \beta})
      ({\bar c}_{L \beta} \gamma_\mu c_{L \alpha}) \,,\nonumber \\
O_3&=&({\bar s}_{L \alpha} \gamma^\mu b_{L \alpha})
      [({\bar u}_{L \beta} \gamma_\mu u_{L \beta})+...+
      ({\bar b}_{L \beta} \gamma_\mu b_{L \beta})] \,,\nonumber \\
O_4&=&({\bar s}_{L \alpha} \gamma^\mu b_{L \beta})
      [({\bar u}_{L \beta} \gamma_\mu u_{L \alpha})+...+
      ({\bar b}_{L \beta} \gamma_\mu b_{L \alpha})] \,,\nonumber \\
O_5&=&({\bar s}_{L \alpha} \gamma^\mu b_{L \alpha})
      [({\bar u}_{R \beta} \gamma_\mu u_{R \beta})+...+
      ({\bar b}_{R \beta} \gamma_\mu b_{R \beta})] \,,\nonumber \\
O_6&=&({\bar s}_{L \alpha} \gamma^\mu b_{L \beta})
      [({\bar u}_{R \beta} \gamma_\mu u_{R \alpha})+...+
      ({\bar b}_{R \beta} \gamma_\mu b_{R \alpha})] \,,\nonumber \\
O_7&=&{e \over 16 \pi^2} \left ( m_b {\bar s}_{L \alpha} \sigma^{\mu \nu}
     b_{R \alpha} +m_s {\bar s}_{R \alpha} \sigma^{\mu \nu}
     b_{L \alpha}  \right) F_{\mu \nu}\,, \nonumber \\
O_8&=&{g_s \over 16 \pi^2} m_b \Big[{\bar s}_{L \alpha}
\sigma^{\mu \nu}
      \Big({\lambda^a \over 2}\Big)_{\alpha \beta} b_{R \beta}\Big] \;
      G^a_{\mu \nu} \,,\nonumber\\
O_9&=&{e^2 \over 16 \pi^2}  ({\bar s}_{L \alpha} \gamma^\mu
     b_{L \alpha}) \; {\bar \ell} \gamma_\mu \ell \,,\nonumber \\
O_{10}&=&{e^2 \over 16 \pi^2}  ({\bar s}_{L \alpha} \gamma^\mu
     b_{L \alpha}) \; {\bar \ell} \gamma_\mu \gamma_5 \ell \,,
\label{eff}
\end{eqnarray}
\noindent with $\alpha$, $\beta$  color indices,
$\displaystyle b_{R,L}={1 \pm \gamma_5 \over 2}b$, and
$\displaystyle \sigma^{\mu \nu}={i \over
2}[\gamma^\mu,\gamma^\nu]$; $e$ and $g_s$ are the electromagnetic
and the strong coupling constant, respectively,  and $F_{\mu
\nu}$ and $G^a_{\mu \nu}$ in $O_7$ and $O_8$ denote the
electromagnetic and the gluonic field strength tensor. $O_1$ and
$O_2$ are current-current operators, $O_3,...,O_6$   QCD
penguin operators, $O_7$  and $O_8$  magnetic penguin operators, $O_9$ and
$O_{10}$  semileptonic electroweak penguin operators. The
Wilson coefficients  in (\ref{hamil}) are known
at NNLO in the Standard Model  \cite{nnlo}.  The operators $O_1$ and $O_2$  contribute to the the final state with a lepton pair
through a $\bar c c$ contribution that can give rise to charmonium resonances $J/\psi$, $\psi(2S)$, $\cdots$,
 resonant term which can be subtracted by appropriate
kinematical cuts around the  resonance masses. The Wilson coefficients $C_3-C_6$ are small,  hence  the contribution of only the operators $O_7$, $O_9$ and $O_{10}$
can be kept for the description of the  $b \to s \ell^+  \ell^-$ transition. In our study we use  a modification of the Wilson coefficient $C_7$:  $C_7^{eff}$, which is a renormalization
scheme independent combination of $C_7, C_8$ and $C_2$, given by a
 formula that  can be found, e.g., in \cite{Colangelo:2006vm}.

The $\bar B_s$ and $f_0$  matrix elements of the operators in (\ref{eff}) can be written in terms of form factors, so that   the differential decay width of $\bar B_s\to f_0\ell^+\ell^-$ reads:

 \begin{eqnarray}
 &&\frac{d\Gamma(\bar B_s\to f_0\ell^+\ell^-)}{dq^2}= \nn \\
&& \frac{G_F^2 \alpha^2_{em} |V_{tb}|^2|V^*_{ts}|^2 \sqrt{\lambda}}{512 m_{B_s}^3\pi^5}
 \sqrt{\frac{q^2-4m_\ell^2}{q^2}}\frac{1}{3q^2} \nn \\ &&\times\Bigg[6m_\ell^2 |C_{10}|^2(m_{B_s}^2-m_{f_0}^2)^2F_0^2(q^2) \\
 &&+(q^2+2m_\ell^2)\lambda\bigg| C_9F_1(q^2)+\frac{2C_7(m_b-m_s) F_T(q^2)}{m_{B_s}+m_{f_0}}\bigg|^2  \nonumber\\
 &&+ |C_{10}|^2(q^2-4m_\ell^2)\lambda F_1^2(q^2)
\Bigg]\nonumber,\label{eq:partial-decay}
 \end{eqnarray}
 with
$\lambda=\lambda(m_{B_s}^2,m_{f_0}^2,q^2)=(m_{B_s}^2-q^2-m_{f_0}^2)^2-4m_{f_0}^2q^2$,
$\alpha_{em}=1/137$  the fine structure constant and $m_\ell$ 
the lepton mass.

Analogously, the SM effective Hamiltonian  for  $b \to s \nu \bar \nu$,
\begin{equation}
H_{b \to s\nu \bar \nu}= {G_F \over \sqrt{2}} {\alpha_{em} \over 2 \pi
\sin^2(\theta_W)} V_{tb} V_{ts}^* \eta_X X(x_t) \, O_L  \equiv C_L
O_L \,\,, \label{hamilnu}
\end{equation}
 includes the operator
\begin{equation}
O_L = \left( {\bar s}\gamma^\mu (1-\gamma_5) b \right) \left({\bar \nu}\gamma_\mu
(1-\gamma_5) \nu\right) \,\, .\label{opnu}
\end{equation}
 $\theta_W$  is the Weinberg angle; the  function $X(x_t)$
($x_t=\displaystyle{ m_t^2 \over m_W^2}$,  with $m_t$  the top
quark mass and $m_W$ the $W$ mass) has been computed in \cite{inami}  and
\cite{buchalla,urban}, while  the QCD factor $\eta_X$ is close  to one
\cite{buchalla,urban,Buchalla:1998ba}, so that one can use $\eta_X=1$.
From this effective Hamiltonian,   the differential decay width
\begin{eqnarray}
 \frac{d\Gamma(\bar B_s\to f_0\nu\bar\nu)}{dq^2}&=&3
 \frac{|C_L|^2\lambda^{3/2}(m_{B_s}^2,m_{f_0}^2,q^2)}{96 m_{B_s}^3 \pi^3}|F_1(q^2)|^2\,\, \nonumber
\end{eqnarray}
can be obtained. 

In the numerical calculation  we use 
\begin{eqnarray}
 C_7=-0.30137,\;\;\; C_9&=&4.1696,\;\;\; C_{10}=-4.46418,\nonumber \\ C_L&=&2.62 \times 10^{-9},
\end{eqnarray}
together with  $V_{ts}=0.0387$ and
$V_{tb}=0.999$~\cite{Amsler:2008zzb}.
Using these inputs and $\tau(B_s)=1.47$ ps \cite{Amsler:2008zzb} we find:
\begin{eqnarray}
 {\cal BR}(\bar B_s\to f_0\ell^+\ell^-)&=&  (9.5^{+3.1}_{-2.6})\times
 10^{-8} \nn \\
 {\cal BR}(\bar B_s\to f_0\tau^+\tau^-)&=&(1.1^{+0.4}_{-0.3})\times 10^{-8}\\
 {\cal BR}(\bar B_s\to f_0\nu\bar\nu)&=&(
  8.7^{+2.8}_{-2.4})\times 10^{-7} \,\,\,  \nn
\end{eqnarray}
with $\ell=e,\mu$.
  Our estimate of the NLO effects in the form factors  modifies   the branching ratios to
$ {\cal BR}(\bar B_s\to f_0\ell^+\ell^-)=  (16.7\pm6.1)\times 10^{-8},$
 ${\cal BR}(\bar B_s\to f_0\tau^+\tau^-)=(2.7\pm1.3)\times 10^{-8},$ and
$ {\cal BR}(\bar B_s\to f_0\nu\bar\nu)= (  15.2\pm5.6)\times 10^{-7}.$
These decay modes are therefore accessible at the  LHCb experiment at the CERN Large Hadron Collider
and at a  Super B factory operating at the $\Upsilon(5S)$ peak.

\subsection{Nonleptonic $B_s\to J/\psi f_0$ transition}\label{PA-B}

The study of CP violation and the measurement of the CKM angles mainly proceed through the measurement  of nonleptonic decay modes.  In the $B_s$ sector,  the channel $B_s \to J/\psi \phi$ is  the golden mode to investigate CP violation, and from the analysis of this mode the CDF \cite{Aaltonen:2007he} and D0 \cite{:2008fj} Collaborations at the Fermilab Tevatron  have  obtained values of the $B_s$ mixing phase $\phi_s=-2\beta_s$ much larger than predicted in the SM,  modulo a  large experimental uncertainty.   If confirmed, this measurement would indicate  physics beyond SM.  It is of prime importance  to consider other processes allowing to access  $\beta_s$, namely  $B_s \to J/\psi \eta,  \, J/\psi \eta^\prime$ and $ J/\psi f_0(980)$ in which   the final state is a CP eigenstate  and no angular analysis is required to disentangle  the various CP components,  as needed for  $B_s \to J/\psi \phi$.
However, the reconstruction of $B_s$ modes into $\eta$ and $\eta^\prime$ is experimentally challenging, since the subsequent $\eta$ or $\eta^\prime$ decays involve photons in the final state. The case of $f_0$ seems  feasible, since $f_0(980)$ essentially decays to $\pi^+ \pi^-$ and to $2 \pi^0$ (the decay to $K^+ K^-$ has also been seen) \cite{Amsler:2008zzb}.  Theoretical predictions of $B_s \to J/\psi f_0(980)$ are therefore of great importance.

The quantitative description of nonleptonic decays is  very challenging. The theoretical framework to study such decays is based on the operator product expansion  and renormalization group methods, which allow to write an effective hamiltonian as in the case of the modes considered in the previous section. However, now one has to consider hadronic matrix elements  $\langle J/\psi f_0 |O_i| B_s \rangle$ with $O_i$ four-quark operators, the calculation of which is a nontrivial task. One of the  strategies which has been  exploited is the naive factorization  \cite{Neubert:1997uc},
 in which such  quantities are replaced by products of matrix elements of the weak currents appearing in each one of the operators of the effective hamiltonian relative to the considered process. These  objects are expressed in terms of  meson decay constants and hadronic form factors.
This procedure is affected by several drawbacks,  and various refinements  have been proposed. It has been shown that a theoretical justification of naive factorization in the case of $B$ decays can be found in the heavy quark limit $m_b \to \infty$ only in a limited class of processes  \cite{Beneke:2000ry}. One can  consider the so called generalized factorization approach, in which the Wilson coefficients (or appropriate combinations of them) appearing in the factorized amplitudes are regarded as effective parameters to be fixed from experiment, a procedure  adopted in the following.

Using the factorization ansatz, the decay amplitude  $\bar B_s(p_{B_s})\to J/\psi(p_\psi,\epsilon) f_0(p_{f_0})$ ($\epsilon$ being the $J/\psi$ polarization vector, $p_{B_s}$, $p_\psi$, $p_{f_0}$ the momenta of the three particles)
is given as
\begin{eqnarray}
 {\cal A}(\bar B_s\to J/\psi f_0)&=& \frac{G_F}{\sqrt 2}V_{cb}V_{cs}^* a_2 m_\psi
 f_{J/\psi} \nonumber \\
 &&F_1^{B_s\to f_0}(m_{J/\psi}^2) 2(\epsilon^* \cdot p_{B_s})\,;
\end{eqnarray}
 $f_{J/\psi}$ is the  $J/\psi$ decay constant,   determined from the   $J/\psi\to e^+e^-$ decay width
\cite{Amsler:2008zzb}: $ f_{J/\psi}=(416.3\pm5.3)$ MeV. The factor
$a_2$ is a combination of Wilson coefficients which
 can be extracted from the $B\to J/\psi K$ decays,  under the assumption  that $a_2$ is the same in the two processes. For these decays the branching ratios are known \cite{Amsler:2008zzb}:
\begin{eqnarray}
 {\cal BR}(B^-\to J/\psi K^-)&=& (1.007\pm0.035)\times 10^{-3},\nonumber\\
 {\cal BR}(B^0\to J/\psi K^0)&=& (8.71\pm0.32)\times 10^{-4}.
\end{eqnarray}
In order to extract $a_2$, the form factor $F_1^{B \to K}$ is required.
We use two different parameterizations, obtained by  short-distance (CDSS) \cite{Colangelo:1995jv}  and light-cone QCD sum rules (BZ) \cite{Ball:2004ye}. The result is different for the two sets of form factors, while there is almost no difference whether we use the charged or the neutral channel:
\begin{eqnarray}
 |a_2^{B\to J/\psi K, (CDSS)}|&=&0.394^{+0.053}_{-0.041},\;\;\; \nn \\
 |a_2^{B\to J/\psi K, (BZ)}|&=&0.25\pm 0.03.
\end{eqnarray}
To be conservative with the hadronic uncertainty,  we use
the average value $a_2=0.32\pm 0.11$ of the two values to  compute
${\cal BR}(\bar B_s\to J/\psi f_0)$.
Using $V_{cb}=0.0412$, $V_{cs}=0.997$ \cite{Amsler:2008zzb} and our LO prediction for the $B_s \to f_0$ form factors, we obtain
\begin{eqnarray}
 {\cal BR}(\bar B_s\to J/\psi f_0)=(3.1\pm 2.4)\times
 10^{-4}\label{eq:Bs-Jpsi-f0-LO}
\end{eqnarray}
while, including our estimate of  NLO corrections, the branching fraction is
$ {\cal BR}(\bar B_s\to J/\psi f_0)=(5.3\pm 3.9)\times10^{-4}$.
The rate of  $B_s\to J/\psi f_0$ is large enough to permit a measurement; notice that the branching fraction of $B_s\to J/\psi \phi$ is 
$ {\cal BR}(B_s\to J/\psi \phi)=(1.3\pm 0.4)\times10^{-3}$ \cite{Amsler:2008zzb}.

To gain a better insight on this point,
it is interesting to compare these results to the branching
fraction of $B_s\to J/\psi_L\phi_L$ ($L$ denotes a longitudinally
polarized meson) computed in the factorization approach. Neglecting the mass difference between  $\phi$ and  $f_0$ in the phase space, the ratio of the  branching fractions of the two modes can be written in terms of form factor combinations:\begin{widetext}
\begin{eqnarray}
 R_{f_0/\phi}^{B_s}&=& \frac{{\cal BR}(B_s\to J/\psi f_0)}{{\cal BR}(B_s\to
 J/\psi_L\phi_L)}\nn \\
 &\simeq&
 \frac{[F_1^{B_s\to f_0}(m_\psi^2)]^2 \lambda(m_{B_s}^2,m_\psi^2,m_{f_0}^2)}{\big[A_1^{B_s\to
 \phi}(m_\psi^2)(m_{B_s}+m_\phi){(m_{B_s}^2-m_\psi^2-m_\phi^2)\over 2 m_\phi}-
 A_2^{B_s\to
 \phi}(m_\psi^2){\lambda(m_{B_s}^2,m_\psi^2,m_\phi^2)\over 2 m_\phi (m_{B_s}+m_\phi)}\big]^2}
  = \left\{ \begin{array}{c}
           0.13\pm 0.06  \\
           0.22\pm 0.10 
          \end{array}\right. \label{Rphi}
\end{eqnarray}\end{widetext}
where the two results correspond to the $B_s\to f_0$ form factor
evaluated at the leading order or not. $A_1^{B_s\to \phi}$ $A_2^{B_s\to \phi}$ are  among the $B_s\to \phi$ transition form factors and are taken from
Ref.~\cite{Ball:2004rg}.
In  Ref.~\cite{Stone:2008ak} it was suggested that the ratio  $ R_{f_0/\phi}^{B_s}$ can be inferred from the  ratio of $D_s$ decay widths to $f_0 \pi^+$ and $\phi \pi^+$, obtaining  $R_{f_0/\phi}^{B_s}\simeq 0.2-0.3$, which is compatible with   our  result (\ref{Rphi}).

Another relation has been also proposed  in \cite{Stone:2008ak}  connecting $R_{f_0/\phi}^{B_s}$ to a different observable in $D_s$ decays:
\be 
R_{f_0/\phi}^{B_s}\simeq R_{f_0/\phi}^{D_s}=\displaystyle{{d\Gamma \over dq^2}(D_s^+ \to f_0 e^+ \nu, \, f_0 \to \pi^+ \pi^-) |_{q^2=0} \over {d\Gamma \over dq^2}(D_s^+ \to \phi e^+ \nu, \, \phi \to K^+ K^-) |_{q^2=0}}.
\ee
For this quantity  the CLEO Collaboration has recently provided a measurement: $R_{f_0/\phi}^{D_s}=(0.42 \pm 0.11)$  \cite{:2009fi}  which is larger than our  (\ref{Rphi}).

All the above  considerations show that the mode $B_s \to J/\psi f_0$ must be used, together with the golden mode   $B_s \to J/\psi \phi$, to measure the  $B_s$ mixing phase, mainly because it provides us with a large number of events and  
does not require an angular analysis to separate  different CP components of the final state.  This is also  the case of modes in which  $J/\psi$ is replaced by a spin $0$ charmonium state,  such as  $\chi_{c0}$, modulo the  difficulty of the $\chi_{c0}$ reconstruction.
 $B_s\to \chi_{c0}\phi$ will provide a
side-check when  the number of accumulated data will increase.  Although
$B_s\to \chi_{c0}\phi$ is a suppressed channel in naive factorization, its
branching fraction may not be small due to the intermediate
rescattering mechanism~\cite{Colangelo:2002mj} or because of the contribution of
nonfactorizable diagrams~\cite{Pham:2005ih}  as in $B\to\chi_{c0}K$.  Analogously for  $B_s\to \chi_{c0}\phi$,
 the branching ratio of $B\to \chi_{c0}K^*$ has
been measured ~\cite{:2008hj},
\begin{eqnarray}
 {\cal BR}(\bar B^0\to \chi_{c0}\bar K^{*0})&=&(1.7\pm0.3\pm0.2)\times 10^{-4}, \label{BtoK*chic0neutral} \nn\\
 {\cal BR}(B^-\to \chi_{c0} K^{*-})&=& (1.4\pm0.5\pm0.2)\times 10^{-4}  \nn \\
 &<&2.1\times 10^{-4} \,\,\,\,\, (90\% \,\,CL) \label{BtoK*chic0charged}
\end{eqnarray}
and, on the basis of  $SU(3)_F$ symmetry, the branching fraction of $B_s\to \chi_{c0}\phi$ should be similar.
 
\section{Decay  $D_s\to f_0 e^+ \nu$}

\begin{figure}[b]
\includegraphics[width=0.4\textwidth]{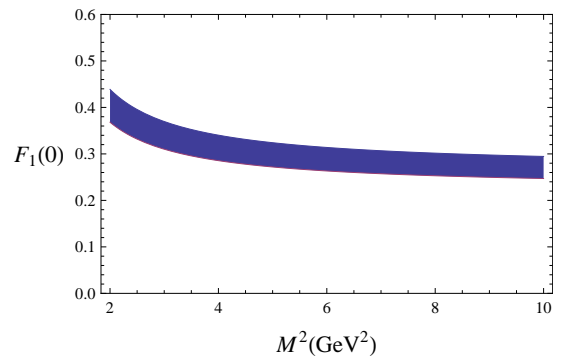}
\caption{Dependence of the $D_s\to f_0$ form factors at $q^2=0$
$F_1(0)=F_0(0)$ on the Borel parameter $M^2$. }\label{fig:M2dependence-Ds}
\end{figure}

\begin{figure}[htth]
\includegraphics[width=0.4\textwidth]{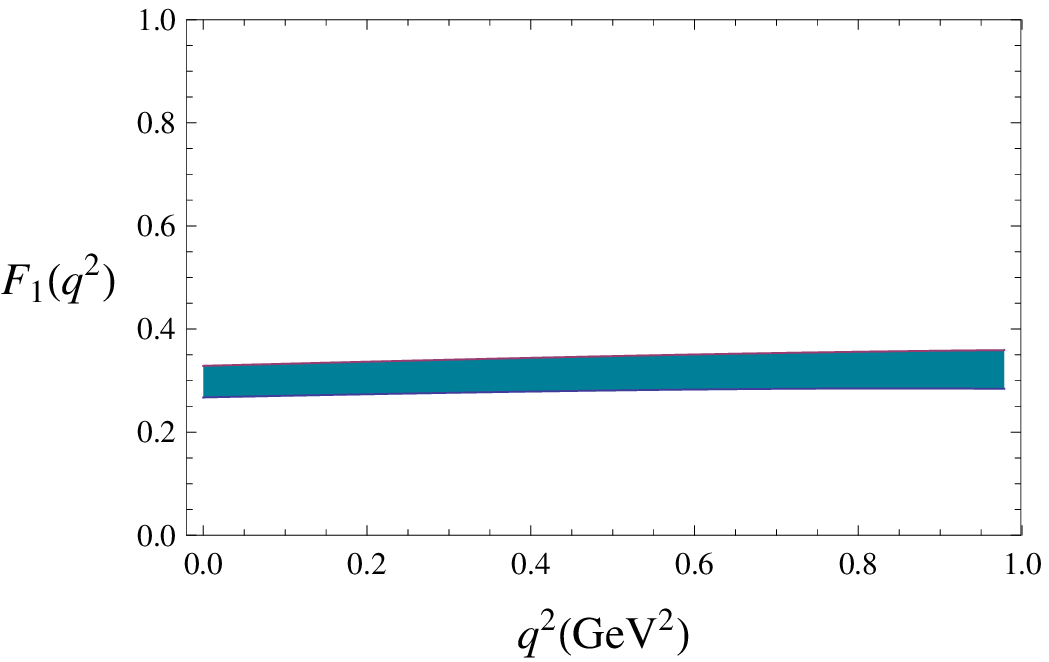}
\includegraphics[width=0.4\textwidth]{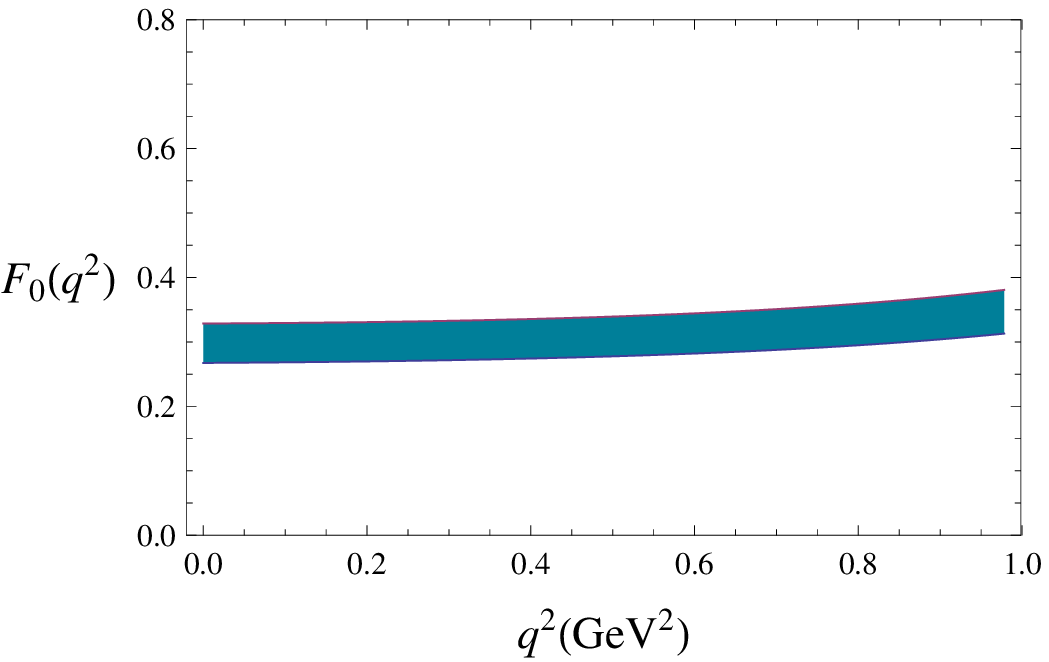} \caption{$q^2$ dependence
of the $D_s\to f_0$ form factors. }\label{fig:LO-formfactor-Ds}
\end{figure}

\begin{figure}[htth]
\includegraphics[width=0.4\textwidth]{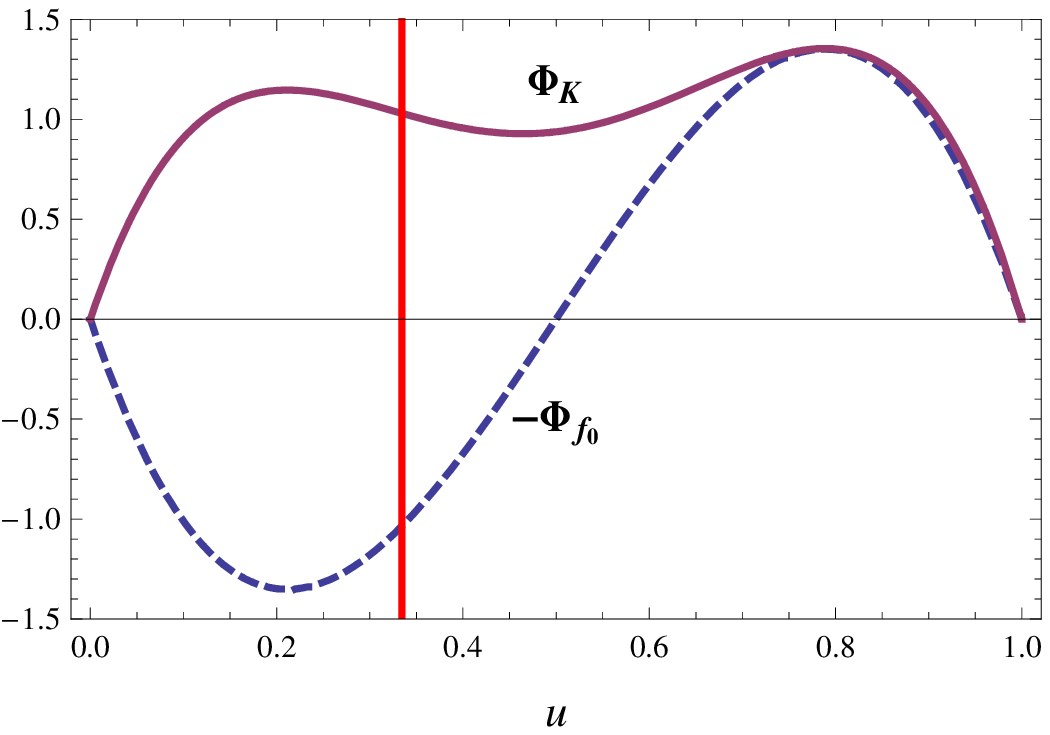} \hskip 0.5cm
\includegraphics[width=0.4\textwidth]{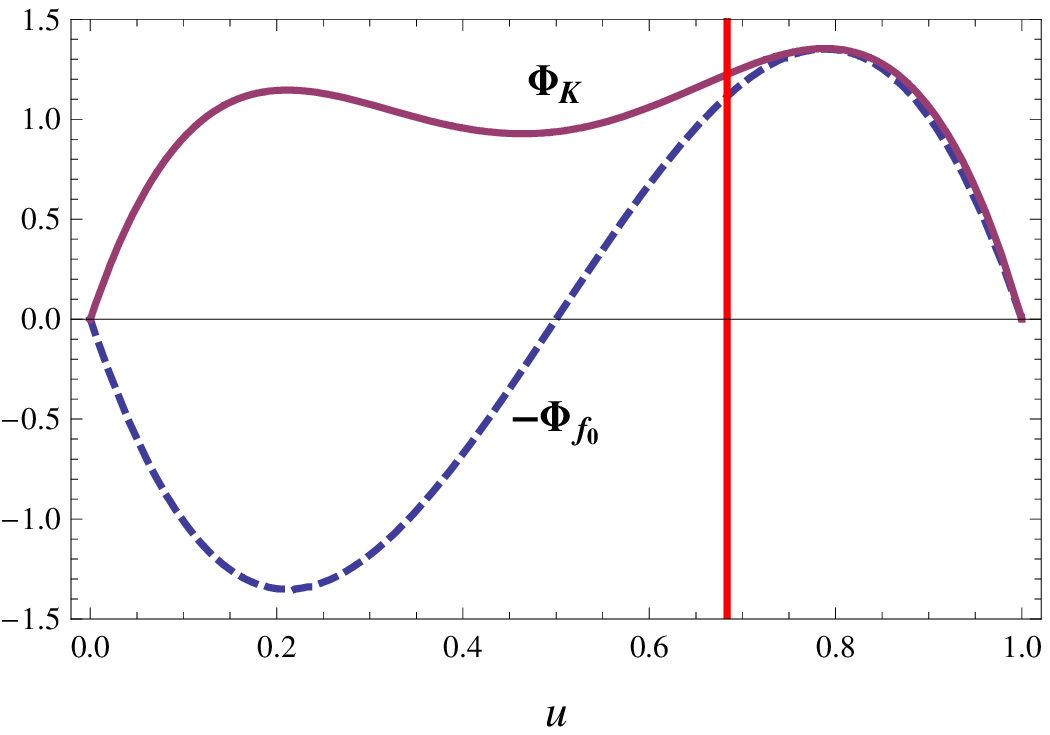}
 \caption{Shape of the twist-2 LCDA:
$-\Phi_{f_0}=-6u(1-u)B_1C_1^{3/2}(2u-1)$ (dashed) and $\Phi_K$
(solid) taken from \cite{ball}.  In the upper panel, the red line denotes the position of
  $u_0^{D_s}=0.334$ fixed for the  $D_s \to f_0$
transition, while in the lower panel the red line corresponds to
 the position of
$u_0^{B_s}=0.684$ at $q^2=0$  in $B_s \to f_0$ transition.
 }\label{fig:T2-LCDA}
\end{figure}

By a suitable change of parameters in the sum rules in Section \ref{sec:formfactors-LCSR},   also the $D_s \to f_0$ form factors can be computed and  the branching ratio of the semileptonic decay $D_s\to f_0 e^+\nu$ can be predicted.  We use $m_c=1.4$  GeV and $\tau(D_s)=0.5$ ps \cite{Amsler:2008zzb}; the threshold parameter is fixed to $s_0^{D_s}=(6.5\pm 1.0)$  GeV$^2$. For the $D_s$ decay constant we use the  value quoted by the
 Heavy  Flavor Averaging Group:  $f_{D_s}=(256.9\pm6.8)$  MeV~\cite{HFAG}.
The Borel parameter can be fixed requiring stability of  the sum rule result with respect to $M^2$ variations.
In Fig.~\ref{fig:M2dependence-Ds} we plot $F_1^{D_s\to f_0}(0)$ versus $M^2$; the stability window is selected in the range $M^2=(5\pm1)$ GeV$^2$. We find
\be
 F_1^{D_s\to f_0}(0)=  F_0^{D_s\to f_0}(0)=0.30 \pm 0.03. \label{F1ds(0)}
\ee
The $q^2$ dependence of the two form factors is displayed  in Fig.~\ref{fig:LO-formfactor-Ds}.
The value of (\ref{F1ds(0)}) is much smaller than in the $D \to K$ case, for which the
light-cone sum rule prediction is: $F_1^{D \to K}(0)=0.75^{+0.11}_{-0.08}$ \cite{Khodjamirian:2009ys}.
We can understand this difference noticing that contribution of the $f_0$
twist-2 LCDA in $D_s\to f_0$ transition is small due to the different shape of the twist-2  $f_0$ distribution amplitude  with respect to the case of $K$. The two LCDA are plotted in Fig.~\ref{fig:T2-LCDA}, where  the position of the parameter $u_0$, defined in Eq.~(\ref{eq:u0}), is also displayed  (upper panel). The situation can be compared to the $B_s \to f_0$ case,  shown in the lower panel of the  figure. Since the LCDA is integrated in the range $[u_0,1]$, one can see that,  in the case of $f_0(980)$, the integral of the distribution amplitude gets two opposite contributions which tend to cancel each other, due to the presence of a zero in the DA. The zero is not  present in the kaon DA, so that the integrated DA gives a much larger contribution. In the case of $B_s$, the position of the parameter $u_0$ is such that the zero of the DA is not included in the integration region, so that no sizable difference is expected between the $f_0$ and the kaon cases.
This argument explains also why, compared with the results of other approaches,
our outcome are smaller. This can be noticed in  Table~\ref{table:Ds-formfactor-comparison}, where
we compare our  results for the $D_s\to f_0$ form factors with other estimates \cite{ElBennich:2008xy,Aliev:2007uu,Ke:2009ed}.
\begin{table}[b]
\caption{$D_s\to f_0(980)$ form factor at $q^2=0$, together with the results obtained by 
CLFD/DR~\cite{ElBennich:2008xy}, 
QCDSR \cite{Aliev:2007uu} and CLFQM~\cite{Ke:2009ed} approaches. }\label{table:Ds-formfactor-comparison}
\begin{tabular}{cccccc}
 \hline \hline&  CLFD/DR &   QCDSR  & CLFQM  & This work \\\hline
  $F_1(0)$ & 0.45/0.46~\footnote{using $f_{D_s}=0.274$ GeV}
  & $1.7(0.27\pm0.02)$~\footnote{using  $f_{D_s}=0.22\pm0.02$ GeV; by using different input parameters two results are obtained, the first one in parentheses, the second one
$1.7$ times larger.}
  & 0.434
  &$0.30\pm0.03$\\
 \hline\hline
\end{tabular}
\end{table}

The form factor $F_1(q^2)$ enters in the expression of the differential decay
 rate 
\begin{eqnarray}
 \frac{d\Gamma(D_s\to f_0e^+\nu)}{dq^2}&=& \frac{G_F^2V_{cs}^2
 \lambda^{3/2}(m_{D_s}^2,m_{f_0}^2,q^2)}{192m_{D_s}^3\pi^3}|F_1(q^2)|^2 \nonumber\\
\end{eqnarray}
where the lepton mass is neglected.  Since   in
$D_s\to f_0e^+\nu$ the kinematically accessible $q^2$ range is limited, the applicable region
for  LCSR is  narrow. One can fit the form factors in the
spacelike region, for example $-2 \, {\rm GeV}^2<q^2<0$, and then
 extrapolate  to the timelike region. However, the result of the extrapolation strongly depends on the choice of the fitting region. Moreover, looking at Fig. \ref{fig:LO-formfactor-Ds}, one can notice that the $q^2$ dependence of  $F_1$ and $F_0$ is  mild. In view of this, we use a constant form factor $F_1(q^2)=F_1(0)$ to compute the branching ratio of $D_s\to f_0e^+\nu$;
the result varies less than 10$\%$ including the $q^2$ dependence according to different fitting formulae.
The obtained  branching fraction is 
\begin{eqnarray}
 {\cal BR}(D_s\to f_0e^+ \nu) =
(2.0^{+0.5}_{-0.4})\times 10^{-3}\;\;  . \label{eq:BR-Dsf0-LCSR}
\end{eqnarray}
The modification due to radiative corrections can be estimated  as in the case of $B_s \to f_0$, finding
$ F_1^{D_s\to f_0}(0)=  0.29^{+0.05}_{-0.04}.$

Let us consider  the available experimental data.
The CLEO Collaboration has measured  the  product of branching fractions \cite{:2009fi}
\begin{eqnarray}
 {\cal BR}(D_s\to f_0(980) e^+ \nu)\times {\cal BR}(f_0\to \pi^+\pi^-)=\nonumber \\
 (0.20\pm0.03\pm0.01)\times 10^{-2},\label{product-ds1}
 \end{eqnarray}
updating a previous determination \cite{:2009cm}
\bea
 {\cal BR}(D_s\to f_0e^+ \nu)\times {\cal BR}(f_0\to \pi^+\pi^-)=\nonumber \\
 (0.13\pm0.04\pm0.01)\times 10^{-2}.\label{product-ds}
\eea
Using experimental data provided by the BES Collaboration
studying the processes $\chi_{c0}\to f_0(980) f_0(980) \to \pi^+\pi^- K^+K^-$ and
$\chi_{c0}\to f_0(980) f_0(980)\to K^+K^-K^+K^-$ ~\cite{Ablikim:2004cg}, CLEO quotes
\begin{eqnarray}
 {\cal BR}(f_0\to \pi^+\pi^-)&=& (50^{+7}_{-9})\times 10^{-2}
\end{eqnarray}
which, combined with  (\ref{product-ds1}),  gives
\be
 {\cal BR}(D_s\to f_0e^+ \nu)=(4.0\pm0.6\pm0.6) \times 10^{-3},
\ee
 marginally consistent with our (\ref{eq:BR-Dsf0-LCSR}).

\section{Conclusions}

We have computed the $B_s\to f_0$ transition form factors using
light-cone QCD sum rules at
 leading order in the strong coupling constant,  and also estimating the size of NLO corrections.  The resulting   form factors permit to predict the rates of   $B_s\to
f_0\ell^+\ell^-$ and $B_s\to f_0\nu\bar\nu$ decays, finding branching ratios  accessible at   future machines, like a Super B
factory,  and  at the LHCb experiment at CERN. 
The branching ratio of $B_s\to J/\psi f_0$ can be predicted
 under the factorization assumption:  we find ${\cal BR}(B_s\to J/\psi f_0)/{\cal BR}(B_s\to
J/\psi \phi)\sim0.2-0.3$,  thus the $B_s\to J/\psi
f_0$ channel can be considered another promising mode to access the $B_s-\bar B_s$ mixing phase.
We have also investigated the $D_s\to f_0e^+\nu$
decay channel by the LCSR approach and compared the results to recent measurements.

\vspace*{0.7cm}
\begin{acknowledgments}
WW  thanks Yu-Ming Wang for useful discussions.
This work was supported in part by the EU Contract No. MRTN-CT-2006-035482, "FLAVIAnet".
\end{acknowledgments}

\end{document}